# TOWARDS THE UNDERSTANDING OF HUMAN DYNAMICS

Tao Zhou, Xiao-Pu Han, and Bing-Hong Wang

Quantitative understanding of human behaviors provides elementary comprehension of the complexity of many human-initiated systems. A basic assumption embedded in the previous analyses on human dynamics is that its temporal statistics are uniform and stationary, which can be properly described by a Poisson process. Accordingly, the interevent time distribution should have an exponential tail. However, recently, this assumption is challenged by the extensive evidence, ranging from communication to entertainment and work patterns, that the human dynamics obeys non-Poisson statistics with heavy-tailed interevent time distribution. This review article summarizes the recent empirical explorations on human activity pattern, as well as the corresponding theoretical models for both task-driven and interest-driven systems. Finally, we outline some future open questions in the studies of the statistical mechanics of human dynamics.

1. **Introduction**

Human behavior, as an academic issue in science, has a history of about one century from Watson [Watson, 1913]. As a joint interest of sociology, psychology and economics, human behavior has been extensively investigated during the last decades. However, due to the complexity and diversity of our behaviors, the in-depth understanding of human activities is still a long-standing challenge thus far. Actually, up to now, most of academic reports on human behaviors are based on clinical records and laboratory data, and most of the corresponding hypotheses and conclusions are only qualitative. Therefore, at least we have to ask two questions: (i) Could those laboratory observations properly reflect the real-life activity pattern of us? (ii) Can we establish a quantitative theory for human behaviors?

Barabási provided a potential start point of those questions [Barabási, 2005a], that is, extracting the statistical laws (especially the scaling laws of temporal statistics) of human behaviors from the historical records of human actions. Traditionally, the individual activity pattern is usually simplified as a completely random point-process, which can be well described by a Poisson process, leading to an exponential interevent time distribution [Haight, 1967]. That is to say, the time difference between two consecutive events should be almost uniform, and the long gap is hardly to be observed. However, recently, both empirical studies and theoretical analyses display us a far different scenario: our activity patterns follow non-Poisson statistics, characterized by bursts of rapidly occurring events separated by long gaps. These new findings have significant scientific and commercial potential. As mentioned by Barabási [Barabási, 2005a], models of human activities are crucial for better resource allocation and pricing plans for telephone companies, to improve inventory and service allocation in both online and "high street" retail.

Barabási and his colleagues have opened a new research direction, namely **Human Dynamics**. Although in its infancy, motivated by both the theoretical and practical significances, the studies of human dynamics attract more and more attention. In this article, we summarize recent progresses on this topic, which may be helpful for the comprehensive understanding of the architecture of complexity [Barabási, 2007]. This article is organized as follows. In the next section, we show the empirical evidence of non-Poisson statistics of human dynamics. In section 3 and section 4, the task-driven and interest-driven models are introduced. Finally, we outline some future open problems in the studies of the statistical mechanics of human dynamics.

## 2. Non-Poisson Statistics of Human Dynamics

As mentioned above, in the prior studies, it is supposed that the temporal statistics of human activities can be described by a Poisson process. That is to say, for a sufficiently small time difference $\Delta t$, the probability that one event (i.e. one action) occurs in the interval $[t, t+\Delta t)$ is independent of the time label $t$, and has an approximately linear correlation with $\Delta t$, as

$$P(t, t+\Delta t) \approx \lambda \Delta t, \quad \lambda > 0, \tag{1}$$

where $\lambda$ is a constant. Under this assumption, the interevent time between two consecutive events, denoted by $\tau$, obeys an exponential distribution as [Haight, 1967]:

$$P(\tau) = \lambda e^{-\lambda \tau}. \tag{2}$$

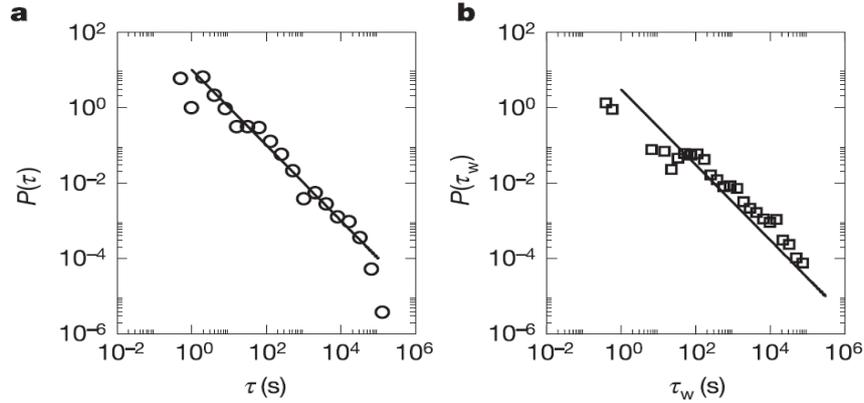

Fig. 1. Heavy-tailed activity patterns in e-mail communications. a, the distribution of the interevent time; b, the distribution of the response time. The data shown in those two plots are extracted from one user. Both the two solid lines have slope -1 in the log-log plot. It is an exact copy from Ref. [Barabási, 2005], and its copyright belongs to the Nature Publishing Group.

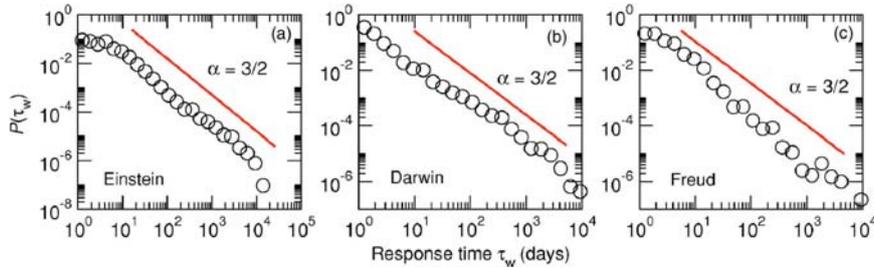

Fig. 2. Distributions of the response times for the letters replied to by Einstein, Darwin, and Freud. All those three distributions are approximated with a power law tail with exponent 1.5. It is an exact copy from Ref. [Vázquez, 2006], and its copyright belongs to the American Physical Society.

Note that, the Eq. (2) decays exponentially, thus a long gap without any event (i.e. a very large $\tau$) should be rarely observed. However, in this section, we show extensive empirical evidence in real human-initiated systems, where the distributions of interevent time have much heavier tails than the ones predicted by Poisson process. Hereinafter, we review the empirical results for different systems separately.

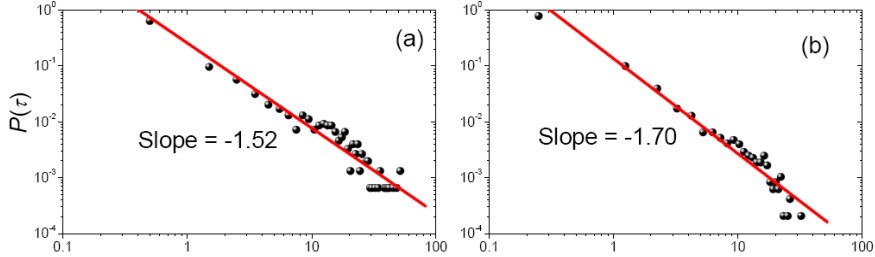

Fig. 3. The distributions of time intervals of sending short-messages on log-log plots, the X-axis denotes time interval (hour), and the Y-axis denotes the probability. Plots (a) and (b) show two typical examples. All those two distributions are approximated with a power law tail with exponent 1.52 and 1.70, respectively. It is an exact copy from Ref. [Hong, 2008].

**E-mail communications**. The dataset contains the email exchange between individuals in a university environment for three months [Eckmann, 2004]. There are in total 3188 users and 129135 emails with second resolution. Denote by $\tau$ the interevent time between two consecutive emails sent by the same user, and $\tau_w$ the response time taking for a user to reply a received e-mail. As shown in Fig. 1 [Barabási, 2005], both the distributions of interevent time and response time obey a power-law form with exponent approximated to 1. Although the exponent differs slightly from user to user, it is typically centered around 1.

**Surface mail communicatons**. The dataset used for analysis contains the correspondence records of three great scientists: Einstein, Darwin and Freud. The sent/received numbers of letters for those three scientists are 14512/16289 (Einstein), 7591/6530 (Darwin), and 3183/2675 (Freud). The dataset is naturally incomplete, as not all letters written or received by these scientists were preserved. Yet, assuming that letters are lost at a uniform rate, they should not affect the main statistical characters. As shown in Fig. 2 [Oliveira, 2005; Vázquez, 2006], the distributions of response time follow a power-law form with exponent approximated to 1.5. The readers should be warned that the power-law fitting for *Freud* is not as good as that for *Einstein* or *Darwin*. Recently, we analyze the correspondence pattern of a Chinese scientist (namely *Xuesen Qian*), and find that both the distributions of interevent time and response time follow a power-law form with exponent 2.1 [Li, 2007].

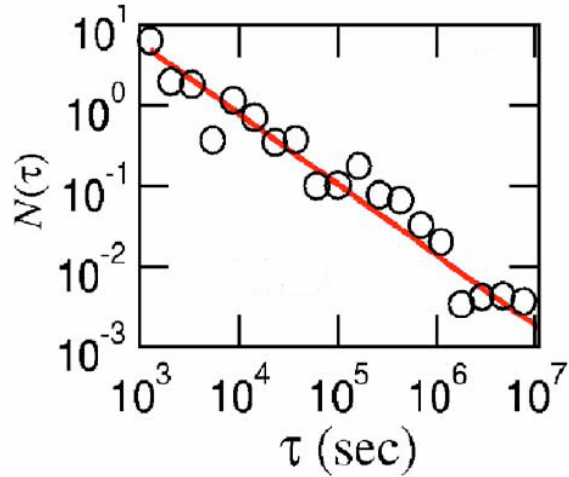

Fig. 4. The distribution of interevent time between two consecutive web visits by a single user. $N(\tau)$ stands for the frequency. The solid line has slope -1.0. It is an exact copy from Ref. [Vázquez, 2006], and its copyright belongs to the American Physical Society.

**Short message communication**. The dataset contained the records of short-message communications of a few volunteers [Hong, 2008]. Figure 3 shows two typical distributions of interevent time between two consecutive short messages sent by an individual. Both the two distributions can be well fitted by power-law functions with different exponents. Actually, in the data resource [Hong, 2008], almost all the distributions of interevent time can be well approximated by power-law forms, and there is an apparently positive correlation between the average numbers of SMs sent per day and the power-law exponents.

**Web browsing**. Browsing history can be automatically recorded by setting cookies. The dataset contains the visiting records of 250000 unique visitors to the site www.origi.hu between Nov. 8[th] to Dec. 8[th] in the year 2002, with about 6500000 HTML hits per day [Dezsö, 2006]. Figure 4 reports the interevent time distribution of a single user, which can be approximately fitted by a power-law function with exponent 1.0. Although the power-law exponents for different users are slightly different, they centered around 1.1. Actually, the exponent distribution obeys a Guassian function with characteristic value approximate to 1.1 [Vázquez, 2006]. Figure 5 reports the interevent time distribution of all the users [Dezsö, 2006], which can be well fitted by a power law with exponent 1.2. Recently, we have analyzed the browsing history through the portal of an Ethernet of a University (University of Shanghai for Science and Technology) in 15 days, with about 4500000 URL requirements per day [Zhao, 2008]. Similar to the above results, we demonstrate the existence of power-law interevent time distribution in both the aggregated and individual levels. However, the exponents, ranged from 2.1 to 3, are much different from the visiting pattern to a single site www.origi.hu.

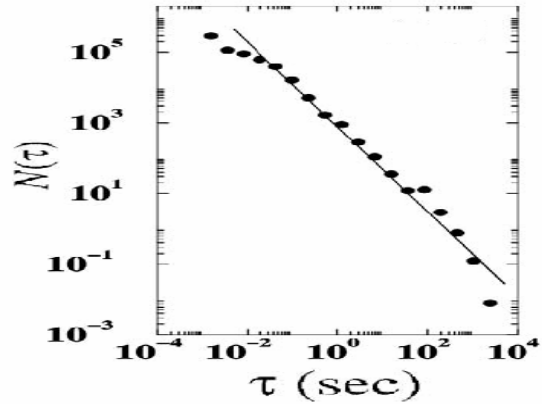

Fig. 5. The distribution of interevent time between two consecutive web visits of users. $N(\tau)$ stands for the frequency. The solid line has slope -1.2. It is an exact copy from Ref. [Dezsö, 2006], and its copyright belongs to the American Physical Society.

**Library loans**. The dataset contains the time books or periodicals were checked out from the library by the faculty at a University (University of Notre Dame) during three years [Vázquez, 2006]. The number of unique individuals is 2247, and the total transaction number is 48409. The interevent time corresponds to the time difference between consecutive books or periodicals checked out by the same patron. Figure 6 reports the interevent time distribution of a typical user, which can be well fitted by a power law with exponent about 1.0. The exponents for different users are different, ranged from 0.5 to 1.5, with the average around 1.0.

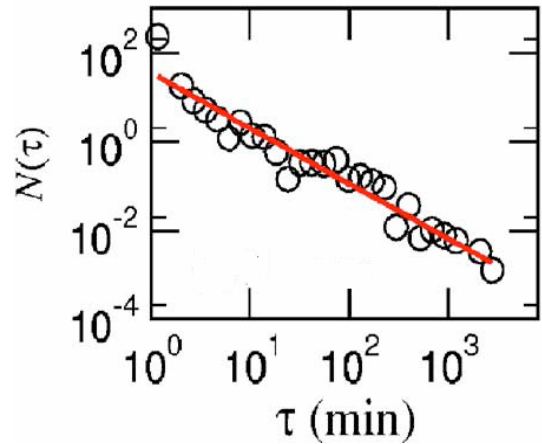

Fig. 6. The distribution of interevent time between two consecutive books or periodicals checked out by the same user. $N(\tau)$ stands for the frequency. The solid line has slope -1.0. It is an exact copy from Ref. [Vázquez, 2006], and its copyright belongs to the American Physical Society.

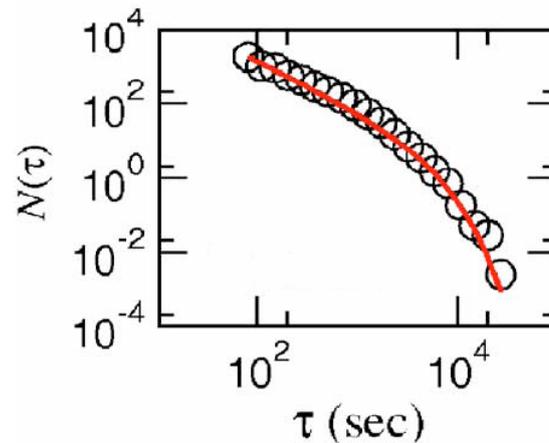

Fig. 7. The distribution of interevent time between two consecutive transactions initiated by a stock broker. $N(\tau)$ stands for the frequency. The solid line represents a truncated power-law form $\tau^{-\alpha}\exp(-\tau/\tau_0)$, where $\alpha=1.3$ and $\tau_0=76$ min. It is an exact copy from Ref. [Vázquez, 2006], and its copyright belongs to the American Physical Society.

**Financial activities**. The dataset contains all buy/sell transactions initiated by a stock broker at a Central European bank between June 1999 and May 2003, with in average ten transactions per day and in total 54374 transactions [Vázquez, 2006]. The interevent time represents the time between two consecutive transactions by the broker, and the gap between the last transaction at the end of one day and the first transaction at the beginning of the next trading day was ignored. Different from the above empirical systems, the interevent time distribution for stock transactions obviously departures from a power-law function. A recent empirical analysis on a double-auction market [Scalas, 2006] also shows that the interevent time between two consecutive orders do not follow the power-law distribution. Although the interevent time distributions in those two financial systems can not be considered as power-law functions, they display clear heavy-tailed behaviors, compared with the exponential form.

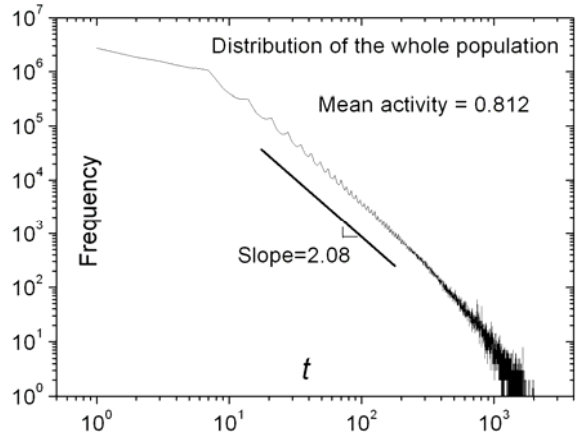

Fig. 8. The distribution of interevent time in the population level, indicating a power-law form. The solid line in the log-log plot has slope -2.08. The data exhibits weekly oscillations, reflecting a weekly periodicity of human behavior, which has also been observed in e-mail communication [Holme, 2003]. It is an exact copy from Ref. [Zhou, 2007a].

**On-line movie watchings**. The data source is collected by a large American company for mail order DVD-rentals, www.netflix.com. The users can rate movies online. In total, the data comprises 17770 movies, 447139 users and about 96.7 millions of records. Tracking the records of a given user *i*, one can get $k_i$-1 interevent times, where $k_i$ is the number of movies *i* has already seen. The time resolution of the data is one day. Figure 8 reports the interevent time distribution based on the aggregated data of all users. The distribution follows a power law for more than two orders of magnitude, with its exponent approximated to 2.08.

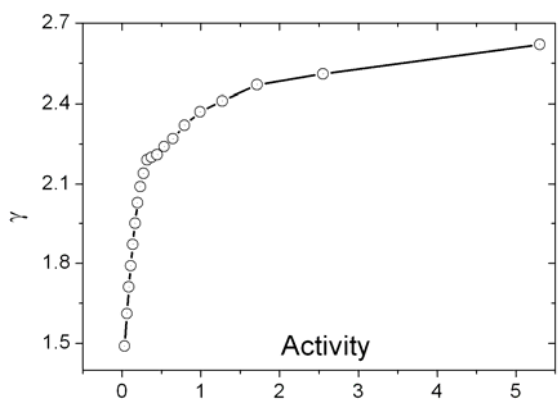

Fig. 9. The relation between power-law exponent $\gamma$ of interevent time distribution and mean activity of each group. Each point corresponds to one group. All the exponents are obtained by using maximum likelihood estimation [Goldstein, 2004] and can pass the Kolmogorov--Smirnov test with threshold quantile 0.9. It is an exact copy from Ref. [Zhou, 2007a].

Although the Poisson processes are widely used to quantify the consequences of human actions, yet, an increasing number of empirical results indicate the non-Poisson statistics of the timing of many human actions, that is, the interevent time $\tau$ or the response time $\tau_w$ obeys a heavy-tailed distribution. Besides what we mentioned above, more evidence could be found in the on-line games [Henderson, 2001], the Internet chat [Dewes, 2003], FTP requests initiated by individual users [Paxson, 1996], timing of printing jobs submitted by users [Harder, 2006], and so on. Actually, the majority of those heavy tails can be well approximated by using a power-low form. However, there also exists the debate about the choice of fitting function for the interevent time distribution in e-mail communication [Stouffer, 2005; Barabási, 2005b]. Another candidate, namely log-normal distribution, has also been suggested [Stouffer, 2005] to be used to describe the non-Poisson temporal statistics of human activities. After the choice of a power-law form, another problem is how to evaluate its exponent [Goldstein, 2004]. In most academic reports, the exponent is directly obtained by using linear fitting in a log-log plot. However, it is found, both numerically and theoretically, that this simply linear fitting will cause remarkable error in evaluating the exponent [Newman, 2005]. A more accurate method, strongly suggested by the authors, is the (logarithmic) maximum likelihood estimation [Goldstein, 2004; Zhou, 2007a].

Based on the analytical solution of the Barabási queuing model [Barabási, 2005], Vázquez *et al.* [Vázquez, 2006] claimed the existence of two universality classes for human dynamics, whose characteristic power-law exponents are 1 and 1.5, respectively. The e-mail communication, web browsing and library loans belong to the former, while the surface mail communication belongs to the latter. However, thus far, there are increasing empirical evidence, as shown above, against the hypothesis of universality classes for human dynamics. For example, we [Zhou, 2007a] sort the Netfilx users by activity (i.e. the frequency of events of an individual, here means the frequency of movie ratings) in a descending order, and then divide this list into twenty groups, each of which has almost the same number of users. As shown in Fig. 9, we observe a monotonous relation between the power-law exponent and the mean activity in the group, which suggests that the activity of individuals is one of the key ingredients determining the distribution of interevent times. And the tunable exponents controlled by a single parameter indicate a far different scenario against the discrete universality classes suggested by Vázquez *et al.* [Vázquez, 2006]. A similar relation between activity and power-law exponent is also reported in the analysis of short message communication [Hong, 2008].

In addition, Vázquez *et al.* [Vázquez, 2006] suggest that the waiting time distribution of the tasks could in fact drive the interevent time distribution, and that the waiting time and the interevent time distributions should decay with the same scaling exponent. Our empirical study on correspondance pattern [Li, 2007] supports this claim. However, the real situation should be more complicated, and a solid conclusion is not yet achieved.

In a word, abundant and in-depth empirical analyses are required before drawing a completely clear picture about the temporal statistics of human-initiated systems.

## 3. Task-Driven Model

What is the underlying mechanism leading to such huamn activity pattern? One potential start point is queuing of tasks. A person needs to face many works in his/her daily life, such as sending e-mail or surface mail, making telephone call, reading papers, writing articles, and so on. Generally speaking, in our daily lives, we are doing these works one by one with some kind of order. In the modeling of human behaviors, we can abstract these activities on human life as tasks. Accordingly, Barabási proposed a model based queuing theory [Barabási, 2005].

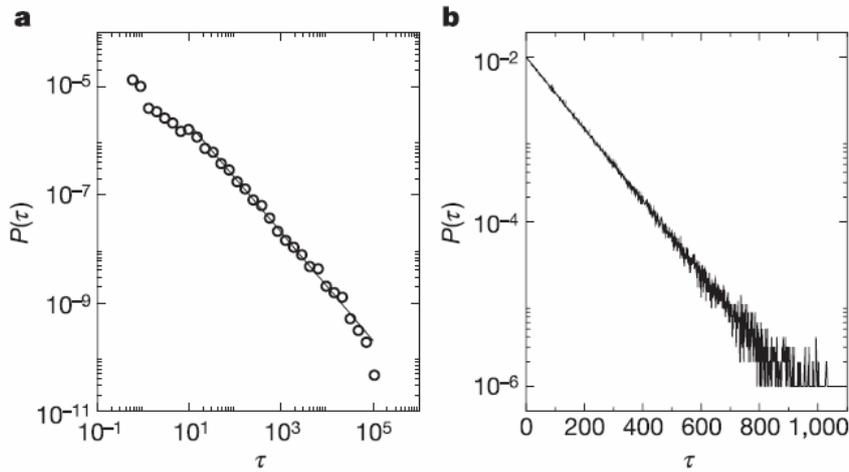

Fig. 10. The waiting time distribution predicted by the queuing model in Ref. [Barabási, 2005]. The priorities were chosen from a uniform distribution $x_i \in [0, 1]$, and the numerical simulation monitores a priority list of length $L = 100$ over $T = 10^6$ time steps. **a**, Log–log plot of the tail of probability $P(\tau)$ that a task spends $t$ time on the list obtained for $p = 0.99999$, corresponding to the deterministic limit of the model. The continuous line of the log–log plot has slope -1, in agreement with the numerical results and the analytical predictions. The data were log-binned, to reduce the uneven statistical fluctuations common in heavy-tailed distributions, a procedure that does not alter the slope of the tail. **b**, Linear-log plot of the $P(\tau)$ distribution for $p = 0.00001$, corresponding to the random choice limit of the model. The fact that the curve follows a straight line on a linear-log plot indicates that $P(\tau)$ decays exponentially. It is an exact copy from Ref. [Barabási, 2005], and its copyright belongs to the Nature Publishing Group.

In this model [Barabási, 2005], an individual is assigned a list with $L$ tasks. The length of list mimics the huamn memory for tasks waiting for execution. At each time step, the individual chooses a task from the list to execute. After being executed, it is removed from the list, and a new task is added. Each task is assigned a priority parameter $x_i$ ($i = 1, 2, \cdots, L$), which is randomly generated by a given distribution function $\eta(x)$. Here the indivudual is facing three possible selection protocols for these tasks:

The first is the first-in-first-out (FIFO) protocol, wherein the individual executes the tasks in the order that they were added to the list. This protocol is common in many service-oriented processes [Reynolds, 2003]. In this case, the waiting time of a task are determined by the cumulative executing time of tasks added into the list before it. If the executing time of each task obeys a bounded distribution, the waiting time, representing the length of time steps between the arrival and execution, of tasks is homogeneous.

The second is to execute the tasks in a random order no matter their priority and arriving time. In this case, the waiting time distribution of tasks is exponential [Gross, 1985].

The last but most important is the highest-priority-first (HPF) protocol. In this case, the tasks with highest priority are executed firstly, even though they are added later in the list. Hence, the tasks with lower priority could wait for a long time before being executed. Such protocol is widely existing in human behaviors: for instance, we usually do the most important or the most urgent works firstly, and then the others.

The Barabási model [Barabási, 2005] focus on the effect of the HPF protocol. At each time step, it assumes that the individual executes the task with the highest priority with probability $p$, and executes a randomly chosen task with probability $1 - p$. Obviously, if $p \to 0$, the model is obey the second protocol, and if $p \to 1$, it displays a pure HPF protocol.

The simulation results with $\eta(x)$ a uniform distribution in [0, 1] are shown in Fig. 10. For $p \to 0$ (random chosen protocol), the waiting time distribution $P(\tau)$ decays exponentially, meanwhile, for $p \to 1$ (HPF protocol), it follows a power-law distribution with exponent 1, which agrees well with the empirical data of e-mail communication. The result shown in Fig. 1(a) are generated with list length $L = 100$, however, the tail of the waiting time distribution $P(\tau)$ is independent with the value of $L$, and the observed heavy-tailed property holds even for $L = 2$. Its exact analysis about the case $L = 2$ for different $p$ is discussed in Refs. [Vázquez, 2005; Vázquez, 2006; Gabrielli, 2007]. As shown in Fig. 11, it is not necessary to have a long priority list for individuals. If an individual can balance at least two tasks, the heavy-tailed property of the waiting time distribution will emerge. These results imply that the HPF protocol could be an important mechanism leading to the non-Poisson statistics of human dynamics.

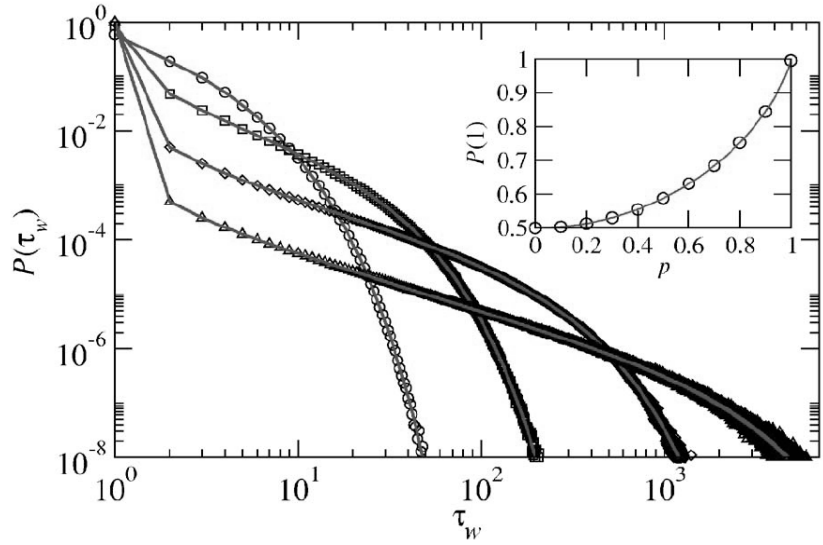

Fig. 11. Waiting time probability distribution function for the Barabási model for $L = 2$ and a uniform new task priority distribution function, $\eta(x) = 1$, in $0 \leq x \leq 1$, as obtained from exact solution (lines) and numerical simulations (symbols), for $p = 0.9$ (squares), $p = 0.99$ (diamonds), and $p = 0.999$ (triangles). The inset shows the fraction of tasks with waiting time $\tau = 1$, namely $P(1)$, as obtained from exact solution (lines) and numerical simulations (symbols). It is a black-white copy from Ref. [Vázquez, 2006], and its copyright belongs to the American Physical Society.

In the further discussions about the Barabási model [Barabási, 2005; Vázquez, 2006], a natural extension is introduced: assuming tasks arrives at rate $\lambda$ and executed at rate $\mu$, and allowing the length of the list of tasks to change in time. Defined $\rho = \lambda/\mu$, obviously, here are three different cases need to be discussed [Vázquez, 2006]:

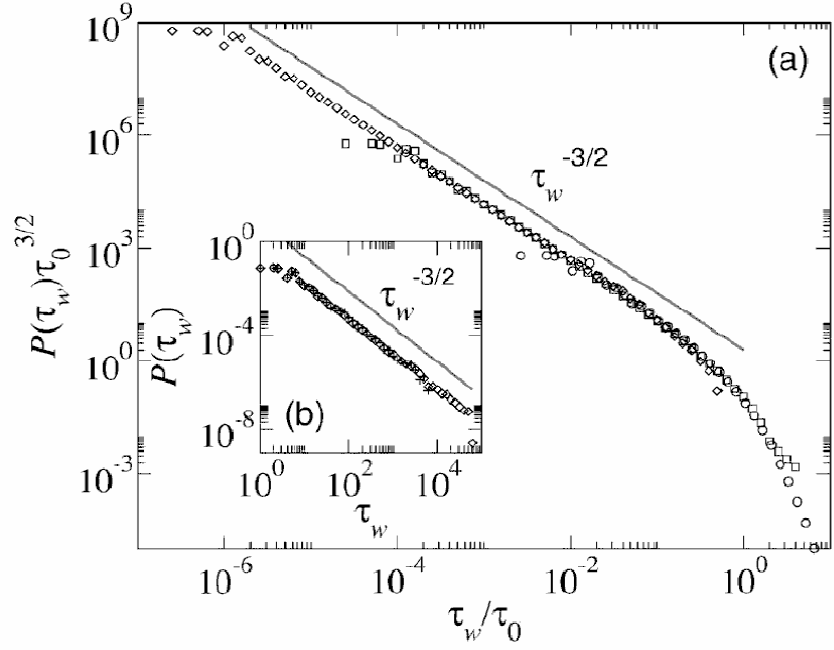

Fig. 12. Waiting time distribution for tasks in the extended queuing model with continuous priorities. The numerical simulations were performed as follows: At each step, the model generates an arrival $\tau_a$ and service time $\tau_s$ from an exponential distribution with rate $\lambda$ and $\mu$, respectively. If $\tau_a < \tau_s$ or there are no tasks in the queue, a new task is then added to the queue, with a priority $x \in [0, 1]$ from uniform distribution, and update the time $t \rightarrow t + \tau_a$. Otherwise, the model removes from the queue the task with the largest priority and update the time $t \rightarrow t + \tau_s$. The waiting time distribution is plotted for three $\rho = \lambda/\mu$ values: $\rho = 0.9$ (circles), $\rho = 0.99$ (squares), and $\rho = 0.999$ (diamonds). The data has been rescaled to emphasize the scaling behavior $P(\tau_w) = \tau_w^{-3/2} f(\tau_w/\tau_0)$, where $\tau_0 \sim (1 - \rho^{1/2})^{-2}$. The inset shows the distribution of waiting times for $\rho = 1.1$, after collecting up to $10^4$ (plus) and $10^5$ (diamonds) *executed* tasks, showing that the distribution of waiting times has a power law tail even for $\rho > 1$ (supercritical regime). Note, however, that in this regime a high fraction of tasks are never executed, staying forever on the priority list whose length increases linearly with time, a fact that is manifested by a shift to the right of the cutoff of the waiting time distribution. It is a black-white copy from Ref. [Vázquez, 2006], and its copyright belongs to the American Physical Society.

The first is called the subcritical regime, where $\rho < 1$, namely the arrival rate is smaller than the execution rate. In this case, the list will be often empty, and most tasks are executed soon after their arrival, thus the long term waiting time is limited. The simulations indicate that the waiting time distribution exhibits an exponential decay when $\rho \to 0$, and when $\rho \to 1$, it is close to a power-law distribution with exponent 3/2 and an exponential cutoff.

The second is the critical regime, where $\rho = 1$. In this case, the length of the list is randomly walking in time. Different from the case with a fixed $L$, the fluctuation in the list length will affect the waiting time distribution. Simulation results indicate that the waiting time distribution obeys a power law with exponent 3/2.

The third is called the supercritical regime, where $\rho > 1$, namely the arrival rate is larger than the execution rate. Thus the length of the list will grow linearly, and a $1 - 1/\rho$ fraction of tasks are never executed. The simulations indicate that the waiting time distribution in this case obeys a power law with exponent 3/2 too. A problem is how to understand such growing list. Let us think over the reply of regular mail. When we received a mail, we put it on desk and piled with early mails, and we usually choose a mail from the pile to reply. If the received mails are too many and we do not have enough time to reply all of them, the mails in the pile will become more and more. We do not need to remember the list, because all the mails are put on there. Therefore, the list of mails waiting for reply is unlimited. The simulation results for this case are in agreement with the empirical data of surface mail replies of *Darwin*, *Einstein* and *Freud*. All of these three great scientists have many mails never being replied.

The numerical results are shown in Fig. 12 [Vázquez, 2006], which is in accordance with the above analysis.

The above model only considers the behavior of an isolated individual. Actually, every person is living surrounding a society with countless interactions to others, such interactions may affect our activities, such as e-mail communication, phone calling and all of the collaborated works. In a recent model on human dynamics, based on the queuing theory, the simplest case taking into account the interactions between only two individuals is considered [Oliveira, 2007]. This model only considers two individuals: A and B. Each individual has two kinds of tasks, interacting task (I) that must be executed in common, and aggregated non-interacting tasks (O) that can be executed by the individual himself/herself. Each task is assigned a random priority $x_{ij}$ ($i =$

I, O; $j$ = A, B) extracted from a probability density function $\eta(x)$. At each time step, both agents select a task with highest priority in their list (with length $L_A$ for A and $L_B$ for B). If both agents select task I, then it is executed, otherwise each agent executes a task of type O.

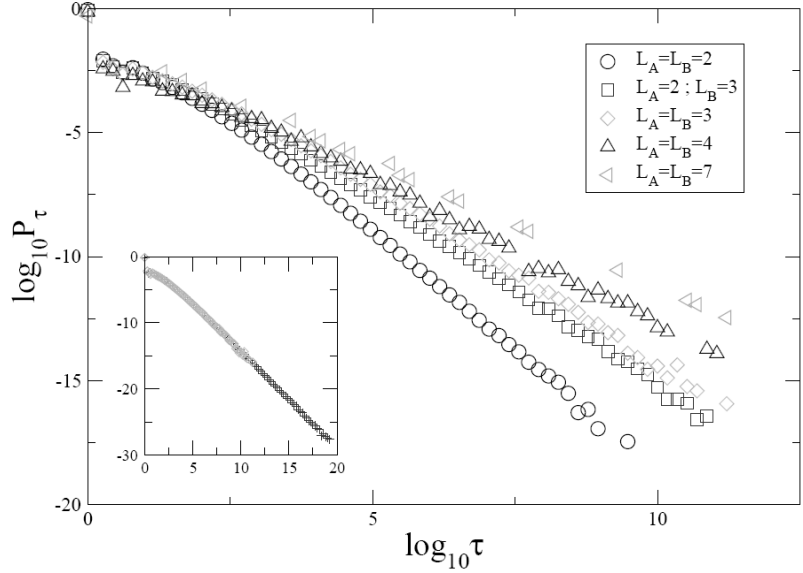

Fig. 13: Probability distribution of the interevent time $\tau$ of the interacting task I, as obtained from the direct numerical simulations of the model. Each dataset was obtained after $10^{11}$ model time steps, corresponding with total number of I plus O task executions. Note that as $L_A$ and/or $L_B$ increases it becomes computationally harder to have a good estimate of $P_\tau$ because the execution of the I task becomes less frequent. It is an exact copy from Ref. [Oliveira, 2007].

As shown in Fig. 13, the numerical simulations of this model indicate that the interevent time distribution of the interacting task I is close to a power law, having a wide range of exponents. This result extends the range of the Barabási model, and highlights a potential way to understand the pattern of the non-Poisson statistics in the interacting activities of human.

**4. Interest-Driven Model and Others**

The motivations of our behaviors are extremely complex [Kentsis, 2006]. Therefore, getting the rule from the simplification or coarse-graining of real world in the modeling is the main way (even the only possible way) for the studies on human dynamics [Oliveira, 2006]. Although the queuing models get a great success in explaining the heavy tails in human dynamics, they have their own limitations. Actually, the core and fundamental assumption of the queuing models is that the behaviors of human are treated as executing tasks, however, this assumption could not fit all the human behaviors. Some real-world human activities could not be explained by a task-based mechanism, but also exhibit the similar statistical law (heavy-tailed interevnt time distribution), such as browsing webs [Dezsö, 2006], watching on-line movies [Zhou, 2007a], playing on-line games [Henderson, 2001], and so on. Clearly, those activities are mainly driven by personal interests, which could not be treated as tasks needing to be executed. The in-depth understanding of the non-Poisson statistics in those interest-driven systems requires a new angle of view beyond the queuing theory [Han, 2007].

Before introducing the rules of an interest-driven model (HZW model for short), let us think over the changing process of our interests or appetites on many daily activities. For example, you had eaten a kind of food with good taste (for example, hamburger) a long time ago. Because so long time you have not tasted it, you almost forgot it. However, after you eat it accidentally, you will remember its good taste, and then your appetite about this food will be stronger and stronger, and the frequency of eaten hamburger gets higher and higher, until you feel you have eaten too much hamburger. Then the good feeling of hamburger disappears, and your appetite is also weakened in a long time. A similar daily experience can also be found in the web browsing. If a person has a long period not browsing the web, an accidental browsing event may give him a good feeling and wake his interest on web browsing. Next, during the activities, the good feeling is durative and the frequency of web browsing may increase. Then, if the frequency is too high, he may worry about it, thus reduces those browsing activities. We can also find the similar changing process of interest or appetite in many other daily activities, such as playing games, watching movies, and so on. In a word, we usually adjust the frequency of the daily activities according to our interest: greater interest will lead to higher frequency, and vice versa. In other words, our interests are adaptively changing.

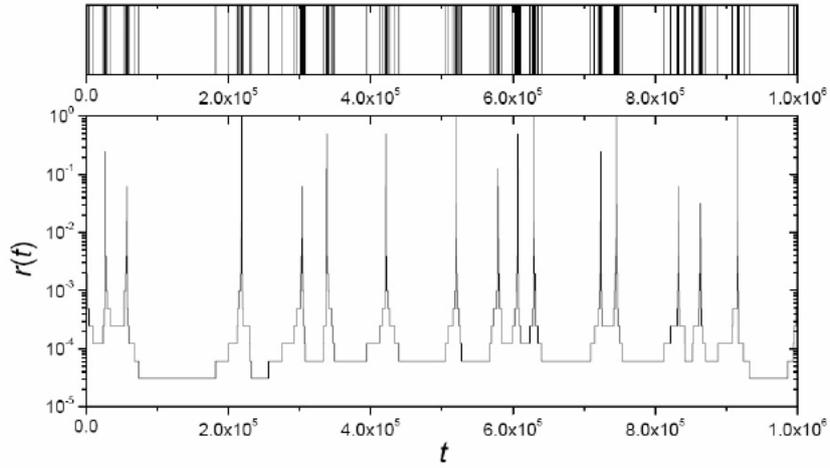

Fig. 14: (upper panel) The succession of events predicted by HZW model. The total number of events shown here is 375 during $10^6$ time steps. (lower panel) The corresponding changes of $r(t)$. The data points are obtained with the parameters $a_0 = 0.5$ and $T_2 = 10^4$. It is an exact copy from the Ref. [Han, 2007].

To mimic these daily experiences, the following simple assumptions in the modeling of the interest-driven system are extracted: Firstly, each activity will change the current interest for a give interest-driven behavior, while the frequency of activities depends on the interest. Secondly, we assume the interevent time $\tau$ has two thresholds: when $\tau$ is too small (i.e., events happen too frequently), the interest will be depressed, thus the interevent time will increase; while if the time gap is too long, we will impose an occurrence to mimic a casual action.

According to above assumptions, the rules of HZW model are listed as follows:

(i)   The time is discrete and labeled by $t = 0, 1, 2, \cdots$, the occurrence probability of an event at time step $t$ is denoted by

       $r(t)$. The time interval between two consecutive events is call the interevent time and denoted by $\tau$.

(ii)      If the $(i+1)^{th}$ event occurred at time step $t$, the value of $r$ is updated as $r(t + 1) = a(t)r(t)$, where $a(t) = a_0$ if $\tau_i \leq T_1$, $a(t) = a_0^{-1}$ if $\tau_i \geq T_2$, and $a(t) = a(t - 1)$ if $T_1 < \tau_i < T_2$.

If no event occurred at time step $t$, we set $a(t) = a(t - 1)$, namely $a(t)$ keeps unchanged. In this definition, $T_1$ and $T_2$ are two thresholds satisfied $T_1 \ll T_2$, and $\tau_i$ denotes the time interval between the $(i+1)^{th}$ and the $i^{th}$ events, and $a_0$ is a parameter controlling the changing rate of occurrence probability ($0 < a_0 < 1$). If no event happens, the value of $r$ will not change. Clearly, simultaneously enlarge (by the same multiple) $T_1$, $T_2$ and the minimal perceptible time, the statistics of this system will not change. Therefore, without lose of generality, the lower boundary $T_1$ is set as 1.

   In the simulations, we set the initial interest, $r_0 = r(t = 0)$, being equal to 1, which is also the maximum value of $r(t)$ in the whole process. As shown in Fig. 14, the succession of events predicted by HZW model exhibits very long inactive periods that separate the bursts of rapidly occurring events, and the corresponding $r(t)$ shows clearly seasonal property. Fig. 15 reports the simulation results with tunable $T_2$ and $a_0$. Given $a_0 = 0.5$, if $T_2 \gg T_1$, the interevent time distribution generated by the present model displays a power law with exponent -1; while if $T_2$ is not sufficiently large, the distribution $P(\tau)$ will departure from the power law, exhibiting a cutoff in the tail. Correspondingly, given sufficiently large $T_2$, the effect of $a_0$ is very slight, thus can be ignored. The power-law exponent -1 can also be analytically obtained under the circumstance $T_2 \gg T_1$, the detailed mathematical derivation can be found in Ref. [Han, 2007].

   Different from the queuing models discussed before, this model is driven by the personal interest. In this model, the frequency of events is determined by the interest, while the interest is simultaneously affected by the occurrence of events. This interplay working mechanism, similar to the active walk [Lam, 2005; Lam 2006], is a genetic origin of complexity of many real-life systems. The rules of the model are extracted from the daily experiences of people, and the simulation results agree with many empirical observations, such as the activities of web browsing [Dezsö, 2006]. This model indicates a much simple activity pattern of human behaviors, that is, a people could adaptively adjust their interest on a specific behavior (e.g. watching TV, browsing web, playing on-line game, etc.), which leads to a quasi-periodic change of interest, and this quasi-periodic property eventually gives raise to the departure of Poisson statistics. This simple activity pattern could be universal in many human behaviors.

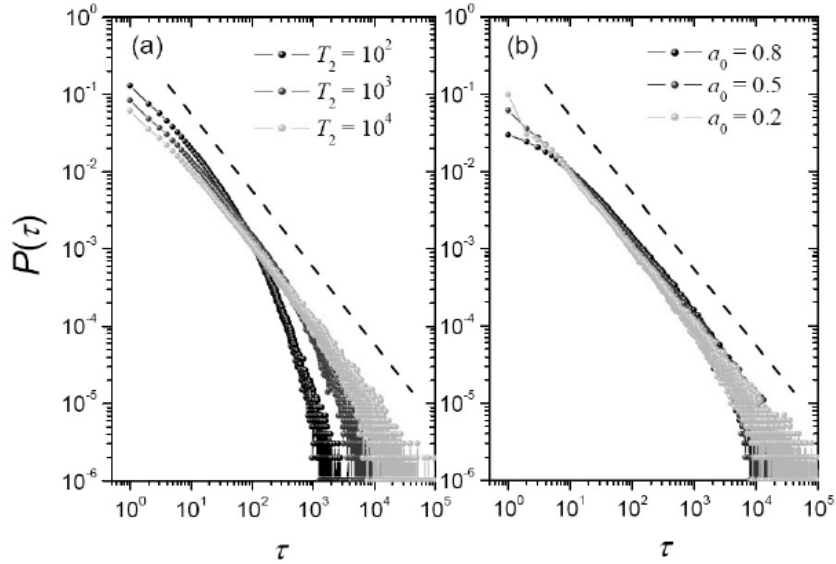

Fig. 15: The interevent time distributions in log-log plots. (a) Given $a_0 = 0.5$, $P(\tau)$ for different $T_2$, where the black, dark gray and bright gray curves denote the cases of $T_2 = 10^2$, $10^3$, and $10^4$, respectively. (b) Given $T_2 = 10^4$, $P(\tau)$ for different $a_0$, where the black, red and green curves denote the cases of $a_0 = 0.8$, 0.5, and 0.2, respectively. The black dash lines in both (a) and (b) have slope -1. All the data points are obtained by averaging over 100 independent runs, and each includes $10^4$ events. It is a black-white copy from the Ref. [Han, 2007].

There are also many other models outside the queuing theory and the interest-driven mechanism. One important model, proposed by Vázquez [Vázquez, 2007a], takes into account the memory of the past activity by assuming that human react by accelerating or reducing their activity rate based on the perception of their past activity rate. Defined $\lambda(t)dt$ is the probability that the individual performs the activity between time *t* and *t*+d*t*. Based on this assumption, the equation of λ(*t*)d*t* can be written as follows:

$$\lambda(t) = a \frac{1}{t} \int_0^t dt' \lambda(t') \quad , \qquad (3)$$

where *a* > 0 is the only parameter in this model. When *a* = 1, λ(*t*) = λ(0) and the process is stationary. On the other hand, when *a* ≠ 1 the process is non-stationary with acceleration (*a* > 1) or reduction (*a* < 1).

Notice, Eq. (3) has a latent assumption of the starting time (*t* = 0). As indicated in Ref. [Vázquez, 2007a], it is a reflection of our bounded memory, meaning that we do not remember or do not consider what took place before that time. For instance, we usually check for new emails every day after arriving at work no matter what we did the day before.

From Eq. (3), we can get the function of interevent time distribution via mathematical derivation. Here we don't introduce the calculations of the model in details; anyone who is interesting in the derivation process can find it from the reference [Vázquez, 2007a]. The general conclusion of the model is as follows:

When *a* = 1, it can generate the interevent time distribution obeying exponential function. Let $\tau_0 = (a\lambda_0)^{-1}$, where $\lambda_0$ is the mean number of events in the considered time period *T*. When *a* > 1 (acceleration) and $\tau_0 \ll \tau < T$, the interevent time distribution generated by the model is close to a power law with exponent $2 + (a - 1)^{-1}$. On the other hand, when 0 < *a* < 1/2 (reduction) and $\tau \ll \tau_0$, its interevent time distribution is close to a power law with exponent $1 - a/(1 - a)$.

Comparing with the cumulative number of regular mails sending as a function of time of *Darwin* and *Einstein*, and the interevent time distribution of these regular mails, the results generated by this simple model are in accordance with the empirical data, as shown in Fig. 16.

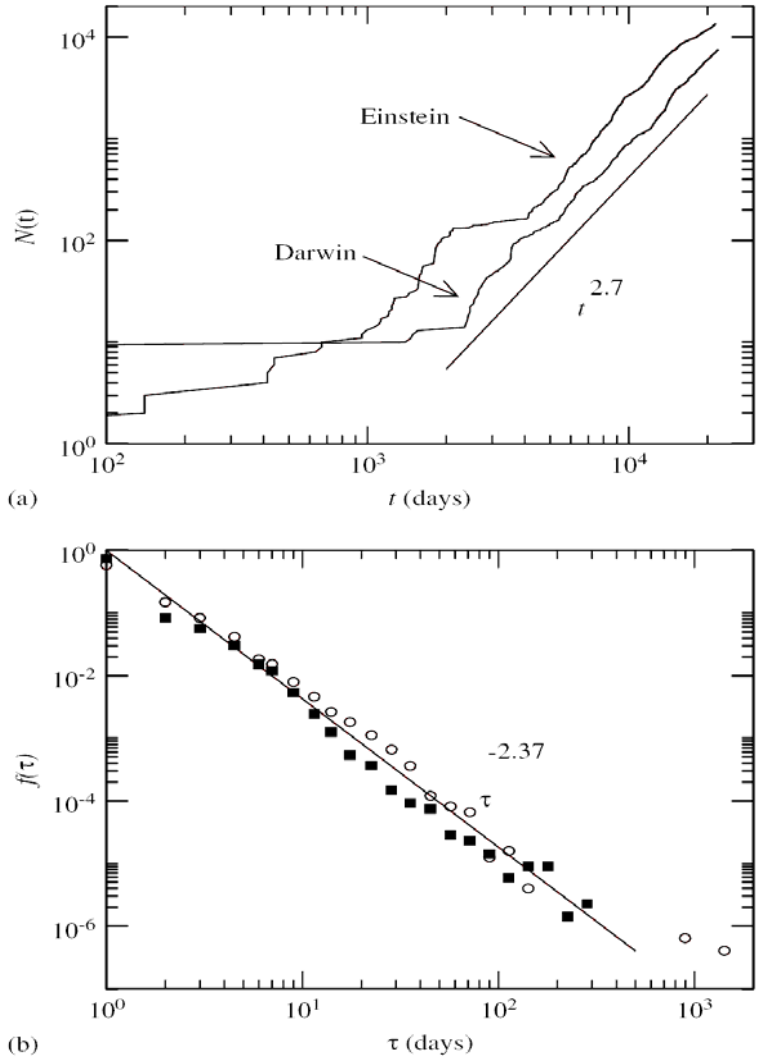

Fig. 16. Regular mail activity: Statistical properties of the Darwin's and Einstein's correspondence. (a) Cumulative number of letters sent by Darwin (open circles) and Einstein (solid squares). The solid line corresponds with a power-law growth. (b) The interevent time distribution associated with the data sets shown in (a). It is an exact copy from Ref. [Vazquez, 2007a], and its copyright belongs to the Elsevier BV.

## 5. Conclusions and Discussions

Although in its infancy, motivated by both the theoretical and practical significances, recently, human dynamics attracts more and more attention. The extensive empirical evidence demonstrates the non-Poisson statistics in temporal human activities, which sharply hit the traditional hypothesis [Gross, 1985] of the uniform and stationary timing of human actions. Actually, the majority of real applications of queueing theory are based on a Poisson process of the event occurrence [Newell, 1982]; hence we may have to complement the queueing theory via taking into account the power-law interevent time distribution. Since the secondary moment of a power-law distribution with exponent smaller than 3 is not convergent, many previous conclusions in queueing theory are not valid when considering a heterogeneous interevent time distribution. Mathematically speaking, the new findings on non-Poisson temporal statistics give rise to a great open question to queueing theory. Besides the theoretical interests, those new findings have significant practical potential [Barabási, 2005a]. For example, the in-depth understanding of human activity pattern is indispensable for the models of social structure [Zhu, 2008] and financial behaviors [Caldarelli, 1997], and is also crucial for better resource allocation and pricing plans for telephone companies [Barabási, 2005a].

Thus far, many models aiming at the explanation of the origin of heavy-tailed human activity pattern are proposed. The mainstream of the prior models are based on queueing theory. However, not all the human-initiated systems are driven by some tasks. Besides the task-driven mechanism underlying the queueing theory, some other possible origins, such as interest [Han, 2007] and memory [Vázquez, 2007a], are also highlighted recently. As stated by Kentsis [Kentsis, 2006], there are countless ingredients affecting the human behaviors, and for most of them, we do not know their impacts. We believe, in the near future, more theoretical models will be proposed to reveal the effects of task deadline, task optimization protocol, human seasonality, social interactions, and so on.

Another important issue is how the non-Poisson temporal statistics affect the relative dynamical processes taking place on the human-initiated systems. For example, the epidemic spreading of diseases, such as AIDS, influenza and SARS, is driven by social contacts between infected and susceptible persons. In the macroscopic level, the effects of social structures (i.e. the epidemic contact networks) have been extensively investigated (see the review article [Zhou, 2006a] and the references therein). However, it lacks a serious consideration of the microscopic factor, namely the temporal statistics of epidemic contacts. Actually, the prior works either assume the contact frequency of a given person proportional to his/her social connectivity [Pastor-Satorras, 2003], or assume an identical contact frequency for all the infected persons [Zhou, 2006b]. Recently, based on the statistical reports of e-mail worms, Vázquez *et al.* [Vázquez, 2007b] studied the impact of non-Poisson activity patterns on epidemic spreading processes, which provides us a starting point of the understanding about the role of individual activity pattern to the aggregated dynamics. In addition, the heterogeneous traffic-load distribution as well as the long-range correlation embedded in the traffic-load time series has been observed in many human-initiated systems, such as the Internet traffic [Park, 2000; ZhouPL, 2006] and air transportation [Guimerà, 2005; Liu, 2007]. The observed heavy-tailed timing of e-mail communication [Barabási, 2005a] and web browsing [Dezsö, 2006], as well as long range human travel [Brockmann, 2006], may contribute to those non-trivial phenomena. Furthermore, some social dynamics may also be highly affected by human activity patterns [Castellano, 2007].

All the previous studies about human dynamics focus on the distributions of interevent time and response time. However, its methodology, exacting statistical laws form the historical records of human activities, is not limited in this issue. For instance, we can also use some of those datasets to quantify the herd behavior of an individual, that is to say, following the opinions of the majority of people in his/her social surrounding in an irrational way. Although the herd behavior has found its significant impact on financial market [Bikhchandani, 2000], it is very hard to be quantified outside the laboratory surrounding [Asch, 1955]. In many web-based recommend systems, such as the on-line movie-sharing system [Zhou, 2007a], the user's records contain not only the time he/she saw the movies, but also his/her opinion (i.e. ratings) on those movies. Similar records can also be found in web-based trading systems, book-sharing systems, music-sharing systems, and so on. Since before the voting on an object, each user can see the previous rating assigned to this object, the herd behavior may occur. Quantitatively uncovering the latent bias of human opinion is crucial for better design of recommender systems [Zhang, 2007; Zhou, 2007b; Zhou, 2007c].

**Acknowledgments**

The authors are benefited from the documents provided by Miss. Nan-Nan Li. We acknowledge Dr. Jian-Guo Liu and Miss. Lin-Yuan Lü for their assistance in the preparation of this manuscript. This work is funded by the National Basic Research Project of China (973 Program No. 2006CB705500), the National Natural Science Foundation of China (Grant Nos. 60744003, 10635040, 10532060 and 10472116), the President Funding of Chinese Academy of Science, and the Specialized Research Fund for the Doctoral Program of Higher Education of China.